\documentclass{Rinton-P9x6}

\begin{document}

\title{Duality in asymmetric quantum optical Ramsey interferometers}

\author{Jes\'us Mart\'{\i}nez-Linares}

\address{Facultad de Ingenier\'{i}a en Tecnolog\'{i}a de la Madera. \\
Edificio D. Ciudad Universitaria. Universidad Michoacana de San Nicol\'{a}s
de Hidalgo. 58060 Morelia, Michoac\'{a}n, M\'{e}xico.\\
E-mail: jesusmartinezlinares@hotmail.com}

\author{Julio Vargas Medina}
\address{Escuela de Ciencias F\'{\i}sico-Matem\'{a}ticas. \\
Edificio B. Ciudad Universitaria. Universidad Michoacana de San
Nicol\'{a}s
de Hidalgo. 58060 Morelia, Michoac\'{a}n, M\'{e}xico.\\
E-mail: \rm{vargasjulio\_2000@yahoo.com.mx}}

\maketitle

\abstracts{ A formalism have been recently derived [J.
Martinez-Linares and D. Harmin, quantum-ph/0306057] allowing one
to separate different sources of which-way information
contributing to the total distinguishability $\mathcal{D}$ of the
ways in a two-way interferometer. Here we apply the formalism to a
Quantum Optical Ramsey Interferometer where both sources, the
\textit{a-priori} predictability of the ways $\mathcal{P}$ and the
quantum "Quality" $\mathcal{Q}$ of the which-way detector, stems
from the same physical interaction. We show that the formalism is
able to separate both sources of which-way information. Moreover,
it is shown that $\mathcal{Q}$ succeeds in quantifying the amount
of quantum which-way information stored in the which-way detector
even in cases where $\mathcal{D}$ does not.}

\section{Introduction}
The duality principle is at the core of the fundamentals of
Quantum Mechanics since its foundations\cite{Feynman65}. In the
last decade,  new results have been found\cite{Scully91} stressing
the role of quantum correlations in the building of quantum
which-way information (WWI) contributing to the distinguishability
of the ways. In fact, quantum correlations can explain the
disappearance of fringes even in situations where the usual
Heisenberg relations cannot\cite{Durr98}. Englert\cite{Englert96}
quantifies the total distinguishability of the ways that can be
potentially available to the experimenter by a parameter
$\mathcal{D}$  and connects it to the fringe visibility
$\mathcal{V}$  measured at the output port of a two-ways
interferometer. Both quantities are related by the inequality
\begin{equation}
\mathcal{D}^2 +\mathcal{V}^2 \le 1, \label{1}
\end{equation}
stating to which degree both distinguishability (particle like
information) and visibility (wave-like information) are
compatible. Equation (\ref{1}) can therefore be interpreted as an
expression of duality.

However, two different sources of WWI are represented in
$\mathcal{D}$. One is the Predictability $\mathcal{P}$ of the
ways, i.e., the \textit{a-priori} WWI that the experimenter has
about the ways. It stems from the preparation of the beam splitter
(BS) and the initial state of the two-level system (Quanton). The
second source is purely quantum-mechanical, stemming from the
quantum correlations established between the Quanton and the WWD,
leading to the storage of WWI in the final state of the which-way
detector (WWD). I an recent work\cite{Jesus03}, from now on cited
as [I], a measure of the latter contribution has been founded.
This measure has been named the "Quality" $\mathcal{Q}$ of the
WWD, since it tell us how "good" the detector is, i.e., it
quantifies the WWD's ability to quantum correlate its final state
with the Quanton's alternatives. Both particle-like information
measures, $\mathcal{P}$ and $\mathcal{Q}$, are related to the
fringe visibility $\mathcal{V}$ through the expression
\begin{equation}
\left( 1-\mathcal{P}^2 \right)\mathcal{Q}^2 +\mathcal{P}^2
+\mathcal{V}^2 \le 1 . \label{2}
\end{equation}
The above equation is also an expression of duality, since it
relates fringe degradation to the availability of different
sources of WWI. In the case we have only one source of WWI, i.e.,
$\mathcal{P}=0$ (or $\mathcal{Q}=0$), we obtain
$\mathcal{D}=\mathcal{Q}$ (or $\mathcal{D}=\mathcal{P}$), and Eq.
(\ref{2}) devolves into Eq. (\ref{1}).

Asymmetric interferometers are not uncommon. A number of proposed
experiments are essentially asymmetric. For instance the Einstein
recoiling slit in a Young double-slit
interferometer\cite{Einstein}, the Quantum optical Ramsey
Interferometer\cite{Ramsey85} (QORI) outlined by Englert
\textit{et al}\cite{Englert92} and the experiment by Haroche
group\cite{newHaroche2001}, in which beam splitting is performed
by the quantized cavity-mode of a high finesse resonator. In all
these cases, both BS and WWD are provided by the same physical
interaction. The asymmetry on such devices is directly coupled to
the ability of the WWD to get entangled with the Quanton. In [I],
we applied the formalism to a quantum logic gate, the Symmetric
Quanton Detecton System (SQDS), in which BS and WWD were
physically independent.  Now, we apply in this paper the formalism
to a QORI. We will show that the formalism is able to separate
both contributions $\mathcal{Q}$ and $\mathcal{P}$ even though
they are both provided the same physical source.

\begin{figure}[t]
\epsfxsize=28pc 
\epsfbox{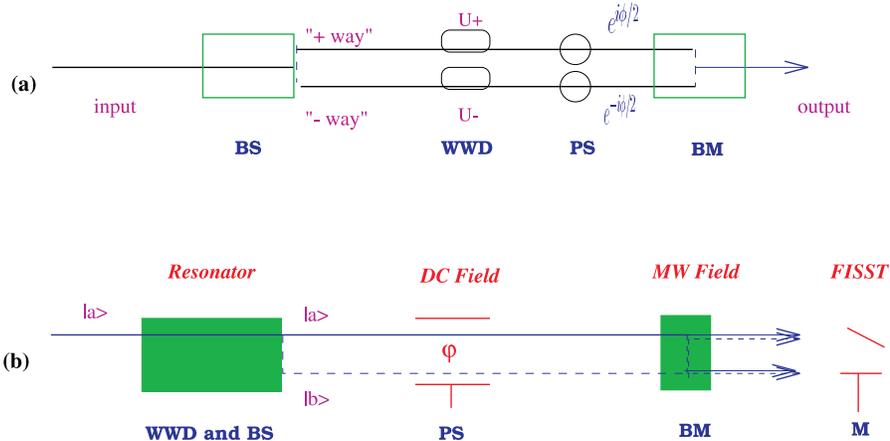} 
\caption{(a)~\noindent Schematic two-way interferometer setup.
(b)~\noindent Quantum Optical Ramsey interferometer. The $\pm
$-ways are realized by electronic states $\{|a\rangle ,|b\rangle
\}$ of a 2-level Rydberg atom. The atomic transition is resonant
with a cavity mode of a high finesse resonator acting as both BS
and WWD. The PS is realized by a dc electric field, which Stark
shifts the atomic levels. The BM is provided by a classical
microwave resonant field performing a $\pi /2$ pulse on the state
of the atom. The final output is measured in {\bf M} by means of a
field ionization state-selective technique  (FISST).
 \label{fig1}}
\end{figure}

\section{The Quantum Optical Ramsey interferometer}

In this section we specialize to the case of the quantum optical
Ramsey interferometer described by Englert \textit{et
al}\cite{Englert92}. A schematic two-way interferometer is
depicted in Fig. \ref{fig1}(a). The beam splitter (BS) distributes
the input states between the 2 ways, which become entangled with
the state of the quantum WWD. The phase shifter (PS) induces a
state dependent shift $\pm \phi /2$. The beam merger (BM)
recombines the contributions into the final state of the quantum.
Measurements of the output build a fringe pattern versus variation
of~$\phi$.

In the case of a QORI\cite{Englert92}, we consider a two-level
atom to be the Quanton and a high finesse resonator to act jointly
as a which-way detector (WWD) and a beam splitter (BS) [see
Fig.\ref{fig1}(b)]. Before entering the resonator, the atom is
prepared in the upper level $|a\rangle$. The atom interacts with
the resonator, adding a photon to its quantized cavity mode if a
resonant transition to the lower level $|b\rangle$ occurs. Due to
the high finesse of the resonator, the cavity field can keep track
of the way taken by the atom since it can store the energy quantum
liberated in the atomic transition. Thus, the same interaction
both splits the beam and makes the two ``ways'' distinguishable.
Next is the turn of the phase shifter (PS)---in the guise, for
example, of an external electrostatic dc field applied at the
central stage of the interferometer\cite{Englert92}. The
differential Stark shift between upper and lower levels induces a
relative phase $\phi$ that can be controlled externally upon
variation of the strength of the applied potential. Finally, a
classical microwave field at the port of the interferometer
supplies the beam merger (BM), effecting a $\pi/2$ pulse after
resonant interaction with the atom. The final state of the atom
after crossing the interferometer is measured by means of
state-selective field ionization techniques at~{\bf M} in Fig.
\ref{fig1}(b). By varying the phase~$\varphi$ in successive
repetitions of the experiment, a fringe pattern can be built up in
the detected probability for the atom to wind up in one state or
the other. It is worth stressing that the atomic center-of-mass
wavefunction is irrelevant for this setup, since it remains
essentially unaffected by the quantum optical Ramsey field crossed
by the atom\cite{Englert94}. Therefore, the two paths refer
exclusively to the internal electronic states of the atom, not to
the actual trajectory of their center of mass.

As commented in the introduction, in this system both the BS and
the WWD are provided by the same physical interaction. Thus,
different preparations of the cavity field in the resonator would
lead to different predictabilities and, in turn, to different
degrees of quantum entanglement with the atoms. Our task is to
quantify the two contributions to the distinguishability: the one
stemming from quantum correlations established after the
interaction between atom and detector, and the other associated
with the asymmetry of the ways resulting from this operation. Both
contributions are related to the fringe visibility at the output
port of the interferometer by the duality expression
\begin{equation}  \label{4.1new}
\left(1-{\cal P}^2\right) {\cal Q}^2 +{\cal P}^2+{\cal V}^2=1 ,
\label{3icssur}
\end{equation}
which invokes Eq.~(\ref{2}) for a system prepared initially in a
pure state\cite{Jesus03}.

Phase-sensitive micromaser schemes have been extensively studied
in the
literature\cite{newHaroche2001,Brune96,Englert95Gantsog,Loeffler96}.
The validity of (\ref{1}) when applied to a QORI has been
demonstrated by Englert\cite{Englert96b}. The analysis of duality
for the QORI is very similar to that presented in [I], once we
remove the unitary constraint on the operators $U_{\pm}$ in Eq.
(5) of [I]. As a matter of fact, choosing for simplicity the
initial pure Quanton state Bloch vector  $\bf{s}_Q^{(0)}=(0,0,1)$,
it is easy to check that Eq. (9) in [I] does give the fringe
visibility contrast factor for the QORI, once the nonunitary
replacements
\begin{eqnarray}
U_+ &\rightarrow& \sqrt{2}\; C^{\dag} ,  \nonumber \\
U_- &\rightarrow& \sqrt{2}\; S a ,  \label{3.1}
\end{eqnarray}
have been taken. Here we have abbreviated the Jaynes-Cummings
operators\cite{Jaynes63,Englert92}
\begin{eqnarray}
S &=& \frac{\sin \left( \Omega \tau \sqrt{aa^{\dag}+\lambda^2}
\right) }{
\sqrt{aa^{\dag}+\lambda^2}}=S^{\dag} ,  \nonumber \\
C &=& \cos \left( \Omega \tau \sqrt{aa^{\dag}+\lambda^2} \right)
+i\lambda S ,  \label{3.2}
\end{eqnarray}
where $a, a^{\dag}$ are the standard photon operators of the cavity mode, $%
\tau$ the interaction time (i.e., the time of flight of the atom
through the resonator), and $\lambda$ is a detuning parameter
normalized to twice the Rabi frequency $\Omega$ of the atomic
transition. In addition, Eq. (17) in [I] is still valid once we
recall that the origin of the asymmetry in the predictability is
the BS action, and not the initial polarization of the Quanton as
was the case in Eqs. (13)--(14) in [I]. The computation of the
relevant quantities of the formalism is now straightforward. In
fact, it suffices to compute
\begin{eqnarray}
w_+ &=& \langle C^{\dag} C \rangle_0 ,  \nonumber \\
w_- &=& \langle S^2 aa^{\dag} \rangle_0 ,  \nonumber \\
{\cal V} &=& 2|\langle S a C \rangle_0| ,  \label{3.3}
\end{eqnarray}
where $w_{\pm}$ are the probabilities that the atom takes the
upper or lower way after the BS, and the averages are taken over
the initial state of the cavity-mode.

\begin{figure}[t]
\epsfxsize=28pc 
\epsfbox{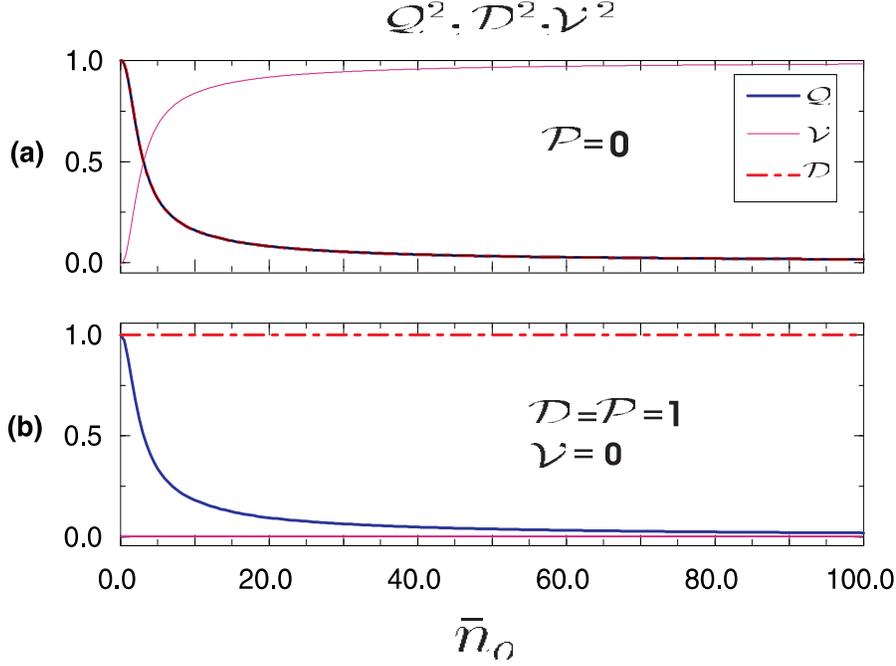} 
\caption[Visibility versus n]{${\cal Q}^2$ (thick solid line),
${\cal D}^2$ (thick dot-dashed line) and ${\cal V}^2$ (thin solid
line) for the quantum optical Ramsey interferometer. The cavity
mode of the resonator is prepared in a coherent state with mean
photon number $\bar{n}_0$. \noindent~(a) Interferometer under
resonant
(detuning $\lambda=0$) and symmetrical (${\cal P}=0$) operation, showing $%
{\cal Q}^2 = {\cal D}^2 = 1 - {\cal V}^2$ (${\cal Q}$ and~${\cal
D}$ coincide, thick lines). The ${\cal P}=0$ condition is achieved
by optimizing
the value of $\Omega\tau$ to assure symmetrical operation of the BS for each~%
$\bar{n}_0$. \noindent~(b) Interferometer under resonant
($\lambda=0$) and maximally asymmetrical  operation (${\cal
P}=1$), so ${\cal V}=0$. In this case, we have optimized the value
of $\Omega\tau$ so that ${\cal P}=1$ is achieved for each
$\bar{n}_0$. Here ${\cal Q}$ and~${\cal D}$ are clearly different:
${\cal D}\ge {\cal P}=1$ saturates ${\cal D}$ to the value of the
predictability. On the other hand, ${\cal Q}$ succeeds in
quantifying the quantum properties of the detector, even in this
case of extreme predictability. } \label{fig2}
\end{figure}

The observation of an interference pattern in the output~{\bf M},
depends on the initial field state $\rho_D^{(0)}$ prepared in the
resonator. If it is unable to acquire which-way information,
quantum interference is observable, due to the
indistinguishability of the paths leading to the same final state.
On the other hand, if the state of the detector becomes perfectly
correlated with the particular path chosen by the atom (for
instance when prepared initially in a Fock state), then no fringes
are observable as implied by duality\cite{Englert92}. In order to
study the transition between these limits, we consider the cavity
mode prepared in a pure coherent state with mean photon number
$\bar{n}_0 \ge 0$ up to large values of $\bar{n}_0$ (e.g., 100).
The numerical results for ${\cal P}^2$, ${\cal Q}^2$, and~${\cal
V}^2$, calculated according to Eqs. (\ref {3.3}) and
(\ref{3icssur}) in this paper, are plotted in Fig. \ref{fig2}. We
also compare the results with the Englert´s distinguishability
satisfying
\begin{equation}
{\cal D}^2 = \left(1-{\cal P} ^2\right){\cal Q}^2 +{\cal
P}^2=1-{\cal V}^2, \label{DQP}
\end{equation}
as shown in Eq. (31) of [I] for the case of pure state
preparation.

%

The symmetrical case is illustrated in Fig.\ref{fig2}(a), where
${\cal Q}^2={\cal D}^2$ (superposed thick lines) and~${\cal V}^2$
(thin line) are plotted at resonance ($\lambda=0$). The values of
the vacuum Rabi phase $\Omega \tau$ have been optimized for each
$\bar{n}_0$ in order to assure symmetrical operation of the BS
(i.e., ${\cal P}=0$). The plot illustrates the loss of coherence
induced by the acquisition of WWI. In the limit of high intensity
of the cavity field, we have $\bf{s}_Q^2={\cal V}^2\rightarrow 1$
and ${\cal Q} \rightarrow 0$. This behavior can be easily
understood, since the pure-state condition and the perfect fringe
visibility that has been reached in this limit are incompatible
with any storage of which-way information in the detector (see
Eq.~(41) in [I] and below). On the other hand, total loss of
coherence is achieved for $\bar{n}_0=0$ (initial vacuum Fock
state). In this case, the detector and the atom evolve into the
maximally entangled state
\begin{equation}
\frac{1}{\sqrt{2}} \left\{ |0\rangle_D |a\rangle_Q + i |1\rangle_D
|b\rangle_Q \right\}.  \label{3.5}
\end{equation}
Perfect correlations are established between atom and detector,
leading to maximum which-way information storage in the cavity
field. Notice that the state (\ref{3.5}), after tracing over the
detector degrees of freedom, describes a totally unpolarized
ensemble, with $\rho_Q=\frac12$ and $\bf{s}_Q^2=0$. We can say
that all the information about the polarization of the atom has
been transferred to the detector.

The difference between ${\cal Q}$ and~${\cal D}$ becomes apparent
in the case ${\cal P}\neq 0$ [see Eq.~(\ref{DQP})]. Actually,
${\cal Q}$ can quantify the amount of which-way information that
can be contained in the detector even in cases where~${\cal D}$
cannot. This is illustrated in Fig.\ref{fig2}(b), where the same
quantities as in Fig.\ref{fig2}(a) have been plotted  for ${\cal
P}=1$. As can be seen in this figure,~${\cal D}$ is tied to the
predictability by the inequality ${\cal D}\ge {\cal
P}=1$\cite{Englert96b}, no matter the degree of which-way
information stored in the state of the detector. This WWI is
however still present in~${\cal Q}$, which keeps invariant with
respect to Fig.\ref{fig2}(a). This property is brought about by
the special structure of the left hand side of relation (\ref{2}),
which cancels the contribution of ${\cal Q}$ (${\cal P}$) in the
case ${\cal P}$ (${\cal Q}$) becomes maximum.

Notice that for $\bar{n}_0=0$ we have both ${\cal Q}=1$ and ${\cal
P}=1$. Full WWI is stored in the quantum state of the cavity-mode
of the resonator even in this extreme asymmetric case ${\cal
P}=1$, that can be associated to an interferometer with a single
way situation ($w_+=1$ or $w_-=1$). This fact can be understood
since the final state of the system (empty cavity $|0\rangle_D
|a\rangle_Q$ or cavity with one photon $|1\rangle_D |b\rangle_Q $)
continues to be  perfectly correlated with the ways taken by the
Quanton (upper or lower way).

\section{Summary}
We consider in this paper a Quantum Optical Ramsey Interferometer
(QORI), for which the Jaynes-Cummings interaction between an atom
and a quantized mode of a cavity field supplies both the beam
splitting and the WWD. We have applied our formalism to the QORI
in order to elucidate the interplay between ${\cal Q}$, ${\cal
P}$, and~${\cal D}$ resulting from this coupling. The value of
${\cal Q}$ can differ substantially from the actual value of
${\cal D}$ in this essentially asymmetric interferometer, since
the latter also involves the \textit{a-priori} which-way
information stemming from the imbalanced beam splitting involved
in the interaction. We have shown that the formalism is able to
separate both contributions. In fact, it is  shown that the
parameter ${\cal Q}$ characterizes solely the quantum properties
of the which-way detector. Moreover, it is shown that~${\cal Q}$
quantifies the amount of which-way information that can be stored
in the detector even in cases where~${\cal D}$ cannot, e.g., when
the latter is saturated by the \textit{a-priori} source of WWI
represented by~${\cal P}$.

The observation of this effect is experimentally feasible. As a
matter of fact, the visibility in Fig. 1(a) have been actually
measured experimentally in the strong-coupling limit,
$\Omega\tau\sim1$, by Bertet {\em et~al}\cite{newHaroche2001} for
cavity field mean photon number ranging between 0 and 14.

%
%


%
%


\section*{Acknowledgments}
J. M.-L. was initially supported  by CIC from Universidad
Michoacana de San Nicol\'{a}s de Hidalgo, and the PROMEP\ program
in M\'{e}xico.

\end{document}